\documentclass[letterpaper,notoc]{JHEP3}
\usepackage[centertags]{amsmath}
\usepackage{array,multirow}
\usepackage{amssymb}
\usepackage{graphicx}
\usepackage{color}

\usepackage{epsfig}
\def\cn{{\mathcal N}}
\def\tr{{\rm Tr}}

\def\o[#1]{{\rm O}\left({#1}\right)}
\def\dotl[#1,#2]{\left\langle #1, #2 \right\rangle}
\def\dotlb[#1,#2]{[ #1, #2 ]}
\def\dotp[#1,#2]{(#1) \cdot (#2)}
\def\>{\rangle}
\def\<{\langle}
\title{Rational Terms in Theories with Matter}
\author{Shailesh Lal and Suvrat Raju \\
Harish-Chandra Research Institute \\
Chatnag Marg, Jhunsi, Allahabad 211019.}
\preprint{HRI/ST/1005}
\abstract{We study rational remainders associated with gluon amplitudes in gauge theories coupled to matter in arbitrary representations. 
We find that these terms
depend on only a small number of invariants of the matter-representation called indices. In particular, rational remainders can depend on the second and fourth
order indices only. Using this, we find an infinite class of non-supersymmetric theories in which rational remainders vanish for gluon amplitudes. This class includes all the ``next-to-simplest'' quantum field theories of \href{http://arxiv.org/abs/0910.0930}{arXiv:0910.0930}. This provides new examples of 
amplitudes in which rational remainders vanish even though
naive power counting would suggest their presence.}
\date{}
\begin{document}
\section{Introduction}
Scattering amplitudes in four dimensional gauge theories have been the subject
of several recent studies \cite{Witten:2003nn}--\nocite{Britto:2004ap,Britto:2005fq,Bern:2005iz,Alday:2007hr,Alday:2007he,Berkovits:2008ic,ArkaniHamed:2008gz,Drummond:2008vq,Drummond:2009fd}\cite{ArkaniHamed:2009dn}. Much of this work has focussed on 
amplitudes in ${\cn=4}$ super-Yang-Mills (SYM) or in pure Yang-Mills 
and has 
involved the development of new techniques to study S-matrix elements in 
these theories. However, these techniques apply far more generally.
 Furthermore, they are capable of shedding fresh light even on familiar and well-studied systems. In this spirit, 
in a previous paper,
we considered gluon scattering amplitudes in gauge theories coupled to matter
in arbitrary representations \cite{Lal:2009gn}.

Using Forde's technique for extracting one-loop integral
coefficients \cite{Forde:2007mi,ArkaniHamed:2008gz} we showed 
that triangle and bubble coefficients in such theories were proportional only to 
a small number of invariants of the matter representation. These 
invariants are called indices.\footnote{The second order index $I_2(R) =  {\tr_{R}\left(T^a T^b\right) \over 2 \tr_F\left(T^a T^b\right)}$ is probably familiar to the reader. As we review later, the trace of a product of any number of generators 
can be expanded in terms of the invariant tensors of the algebra multiplied by coefficients called
indices.  The higher indices are closely related to the higher Casimir invariants.} 
Using this information, 
we were able to find new examples of theories in which gluon scattering 
amplitudes were free of triangles and/or bubbles. 

In this paper, we extend this argument to show that rational terms associated
with gluon amplitudes in theories with matter are also proportional to 
the first few indices (up to the fourth order indices) of the matter representation.\footnote{We should clarify that boxes,
triangles and bubbles come with associated rational terms. In this paper, 
we use the phrase ``rational terms'' to refer to the rational remainders 
that are not associated with these integral functions.} This 
surprising result follows from the newly 
developed method of extracting rational terms by considering the large-mass limit
of massive particles propagating in the loop \cite{Badger:2008cm}.

Rational terms are notoriously difficult to extract since they are missed
by four dimensional unitarity cuts. One has to resort either to $d$-dimensional
unitarity \cite{vanNeerven:1985xr,Bern:1995db,Anastasiou:2006jv,Anastasiou:2006gt,Brandhuber:2005jw,Giele:2008ve} or to other techniques like on-shell recursion at one-loop \cite{Bern:2005hs,Bern:2005cq,Berger:2006ci}.
However, for our purposes the most useful approach is the one developed by 
Badger \cite{Badger:2008cm}. Here, a massless d-dimensional particle propagating
in the loop 
is traded for a massive 4-dimensional particle and rational terms are extracted by 
examining the behaviour of unitarity cuts at large mass.

This approach reveals the remarkably simple structure of rational terms
in gluon amplitudes referred to above.  The fact that the rational contribution
of matter to gluon amplitudes can be written
in terms of the first few indices of the matter representation implies that the 
condition that rational terms vanish can now be expressed in terms of linear 
Diophantine equations involving these indices. We solve these equations to find
an infinite class of non-supersymmetric theories in which rational terms 
vanish for gluon amplitudes. This set 
includes, but is not limited to, the set of next-to-simplest quantum field
theories of \cite{Lal:2009gn}.

This is interesting because these theories are not naively cut-constructible.
Supersymmetric theories are cut-constructible because the expansion of 
an amplitude in terms of Feynman diagrams can be organized to show that 
two powers of the momentum cancel between fermions and bosons \cite{Bern:1994cg,Dixon:1996wi}. In our examples, 
naively 
counting the powers of momentum that appear in Feynman diagrams would lead one
to suspect that rational terms should be present. In this sense the unexpected
simplifications that are present in our theories are similar to those 
seen in ${\cn=8}$ supergravity \cite{BjerrumBohr:2008ji} and QED \cite{Badger:2008rn}.

An overview of this paper is as follows. In section \ref{secreview}, 
we review the results of our previous paper. 
In section \ref{secrationalmatter}, we show that rational terms associated with gluon amplitudes
are proportional to the second and fourth order indices of the matter representation. In section \ref{secdiophantine}, we write down the condition for gluon amplitudes to be
free of rational terms and find new examples of theories in which these are cut-constructible.
We conclude in section \ref{secconclusions}. The appendix contains some
group-theoretic details.

\section{Review}
\label{secreview}
Let us briefly review how triangle and bubble coefficients for gluon amplitudes
in gauge-theories coupled to matter
turned out to be proportional to only a few indices of the matter representation.
 Naively,
we would not expect this at all. For example, consider the following Feynman 
diagram (Fig. \ref{figmatterloop}) for a 10-point gluon amplitude with a massless fermion in the loop.

This 
Feynman diagram is proportional to $\tr(T^{a_1} T^{a_2} \ldots T^{a_{10}})$ where $a_1, \ldots a_{10}$
are colors associated with the gluon lines. So, naively one would certainly
not expect that one-loop integral coefficients for a scattering amplitude
of an arbitrary number of gluons would be sensitive only to the trace of a
small number of generators.
\FIGURE{
\label{figmatterloop}
\epsfig{file=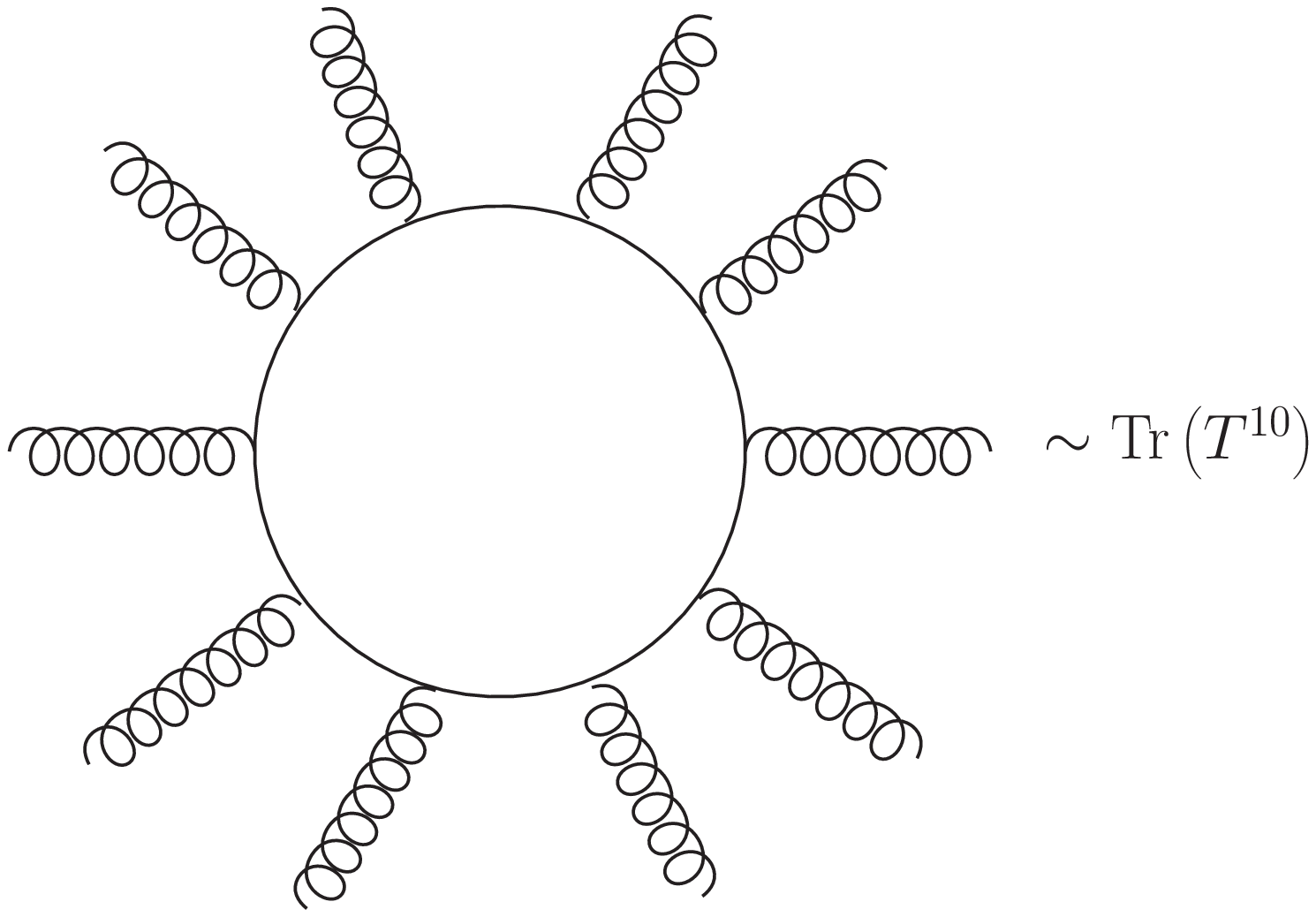, width=0.4\textwidth}
\caption{A Matter-Loop}
}
Of course, we also know that the one-loop $\beta$ function for the gauge-coupling simplifies and is proportional to the quadratic index only. It turns out
the coefficients of triangles and bubbles also simplify similarly. They
are not as simple as the one-loop $\beta$ function and depend on the 
higher-indices also. Triangles can depend on the sixth order indices (these are what appear when we
expand the trace of six generators in terms of the invariant tensors of the
algebra), while bubbles
can depend on the fourth order indices. More precisely, the contribution 
of a scalar or a fermion in representation $R_{s/f}$ to a triangle
coefficient --- $B$--- and a bubble coefficient --- $C$ ---  associated
with a gluon amplitude can be written as (in the notation of 
\cite{Lal:2009gn})
\begin{equation}
\label{trianglebubbleintermsoftraces}
\begin{split}
B_{s/f} &= \sum_{n=2,4,5,6} \omega^B_{a_1 \ldots a_n} \tr_{R_s/R_f}(T^{(a_1} \ldots T^{a_n)}),\\
C_{s/f} &= \sum_{n=2,4} \omega^C_{a_1 \ldots a_n} \tr_{R_s/R_f}(T^{(a_1} \ldots T^{a_n)}), \quad \text{in non-supersymmetric~theories.}
\end{split}
\end{equation}
We emphasize
that this result holds for an {\em arbitrary number} of external gluons.

For supersymmetric theories, these results simplify. For a chiral
multiplet in representation $R_{\chi}$, triangle coefficients
can depend on the higher indices up to the fifth order indices while bubble
coefficients only depend on the quadratic index i.e.
\begin{equation}
\label{susytribubintermsoftraces}
\begin{split}
B_{\chi} &= \sum_{n=2,4,5} \omega^B_{a_1 \ldots a_n} \tr_{R_{\chi}}(T^{(a_1} \ldots T^{a_n)}),\\
C_{\chi} &= \omega^C_{a_1 a_2} I_2(R_{\chi}) \kappa^{a_1 a_2} , \quad \text{in supersymmetric~theories,}
\end{split}
\end{equation}
where $\kappa$ is the Killing form.

This leads to an interesting possibility. Since triangles and bubbles
are sensitive only to a small number of invariants of the representation
and not to the full-character, we can replace the adjoint matter
of the ${\cn=4}$ SYM theory with matter in a different representation that has
 the same first few indices. In this way, one can mimic the adjoint
representation as far as the triangle and bubble coefficients are concerned.

In fact, demanding that the theory be free of triangles and bubbles leads
to linear Diophantine equations involving these higher-order indices. This is
because any representation can be decomposed in terms of irreducible
representations
\begin{equation}
R = \bigoplus n_i R_i.
\end{equation}
Since the indices are linear, mimicking the first few indices of the adjoint
leads to linear equations in the $n_i$ (which are, of course, constrained
to be natural numbers). More precisely, the conditions for supersymmetric
and non-supersymmetric theories to be free of triangles and/or bubbles 
can be written as in Table \ref{simpleconds}.
\TABLE{
\label{simpleconds}
\begin{tabular}{|c|c|c|}\hline
\multicolumn{3}{|l|}{Condition ({\bf C}): ${\rm Tr}_{\rm R}(\Pi_{i=1}^n T^{a_i}) =  m\,{\rm Tr}_{\rm adj}(\Pi_{i=1}^n T^{a_i}),~ n \leq p$} \\ \hline
Non-susy theories have&only boxes & no bubbles \\ \hline
if $R_f$ satisfies {\bf C} with&p=6, m=4&p=4,m=4\\\hline
and $R_s$ satisfies {\bf C} with&p=6, m=6&p=4,m=6.\\\hline
Susy theories have&only boxes & no bubbles \\ \hline
if $R_{\chi}$ satisfies {\bf C} with&p=5, m=3&p=2,m=3.\\\hline
\end{tabular}
\caption{Conditions for the S-matrix to simplify}
}

In our previous paper, we solved these equations. In the planar limit, there
are several theories including the $\cn=2, SU(N)$ theory with $2 N$
fundamental hypermultiplets, in which gluon amplitudes are free
of triangles and bubbles. We found two examples where these
properties persisted even for the non-planar sector.
One of these ---
the ${\cn=2}$ SYM theory with a symmetric and an anti-symmetric tensor hypermultiplet --- is an orientifold of the ${\cn=4}$ theory but the fact that 
its amplitudes at {\em all $N$}  are as simple as those of the ${\cn=4}$ 
theory goes {\em beyond planar equivalence}.

We also found several example of non-supersymmetric theories that were free
of bubbles but had triangles. These theories will make another appearance
below where we show that they are all also free of rational terms.

\section{Rational Terms in Theories with Matter}
\label{secrationalmatter}
We now turn to a study of rational terms associated with gluon amplitudes
in theories coupled to matter in arbitrary representations. As we
review below, gluon
amplitudes in supersymmetric theories are cut-constructible \cite{Bern:1994cg}.
This means that the contribution of fermions, in any representation,
 to rational terms is the same (up to a minus sign) as that of scalars.  Hence, it is sufficient
to consider the contribution of scalars to rational terms. This is what
we do below.

As we mentioned above, our tool
will be the method of extracting rational terms by trading a d-dimensional
massless scalars for a 4 dimensional massive scalar. Rational terms
come from the large-mass limit of massive unitarity cuts. We will find
that the behaviour of tree-amplitudes simplifies in this limit. This means that integral coefficients and the rational terms that they imply
also simplify.

\subsection{Review}
We now quickly review the argument that rational terms vanish in supersymmetric
theories \cite{Bern:1994cg,Dixon:1996wi}. We focus on gluon amplitudes. One-loop gluon amplitudes can be obtained from the 1PI effective action. The 1PI effective action for the gauge field, 
in the
presence of scalars and fermions can be calculated in background field
gauge and written as
\begin{equation}
\label{effectiveaction}
\begin{split}
i \Gamma[A] &= {-i \over 4 g^2} (F_{\mu \nu}^a)^2 + i {\mathcal{L}}_{\text{c.t.}} - {1 \over 2} \ln \text{det}_{[1]} \Delta_{\text{adj},1} + \ln\text{det}_{[0]} \Delta_{\text{adj},0} \\&+ {n_f \over 2} \ln \text{det}_{[1/2]} \Delta_{R_f,{1 \over 2}} - {n_s \over 2} \ln \text{det}_{[0]} \Delta_{R_s,0},
\end{split} 
\end{equation}
where the first two determinants come from the gauge field and ghosts and the next
two come from the fermions and scalars respectively. (See \cite{peskin1995iqf} for a derivation of this result.) For us, it is only important that the generalized d'Alembertians
$\Delta_{r,j}$   have the form
\begin{equation}
\label{dalambertian}
\Delta_{r,j} = -\partial^2 + i\left[\partial^{\mu}A_{\mu}^a T^a_{r} 
+ A_{\mu}^a T^a_{r} \partial^{\mu} \right] + A^{a \mu} T^a_{r} A_{\mu}^b T^b_{r} + F^b_{\rho \sigma} J_{j}^{\rho \sigma} T^b_r.
\end{equation}
where $J_j$ is the generator of Lorentz transformations for spin $j$ and
the  $T^a$
are the generators of gauge transformations for representation $r$.

Now, all one-loop amplitudes can be obtained by attaching tree-graphs to 
the one-loop vertices obtained by expanding these determinants. Consider
 the one-loop integrals that
result from expanding \eqref{effectiveaction} in powers of $A$. Those integrals
that have the same number of momenta in the numerator as propagators 
in the denominator can have no insertion of the last term in \eqref{dalambertian} involving  $F \cdot J$; hence, they cancel
in supersymmetric theories. Furthermore, since $\tr(J_1) = \tr(J_{1 \over 2}) = 0$, we must have at least two insertions of $F \cdot J$.  A loop-integral with two insertions of this term must
have at least two powers of momentum less in the numerator than in the 
denominator. This is enough to ensure that rational terms vanish in 
supersymmetric theories.

Hence, the contribution of scalars, in a certain representation, to rational 
terms is the same as the contribution of fermions in the same 
representation. So, we can obtain all the information we want just by 
considering scalars.

Note, that the formula \eqref{effectiveaction} itself does not tell
us much about the contribution of scalars to rational terms. In fact,
naively expanding the determinants using \eqref{dalambertian} would
lead us to believe that we obtain traces of an arbitrary number
of generators. As we see below, this is not correct.

The contribution of scalars to rational-terms can be conveniently obtained
using the methods of \cite{Badger:2008cm}. We take the scalar propagating
in the loop to be massive, with mass $\mu$, and then pick out specific
coefficients of $\mu$ in the box, triangle and bubble coefficients.

\subsection{Boxes}
The box coefficient is calculated through a product of four tree-amplitudes.
When we consider the contribution of scalars to the box coefficient, each
of these tree-amplitudes has two scalars apart from an arbitrary number of
external gluons. 
According to \cite{Badger:2008cm}, we need to assign mass $\mu$ to
these intermediate scalars and then extract the coefficient of $\mu^4$
in the box-coefficient.

So, consider the coefficient for the box with momenta 
$q_1 \ldots q_4$ at the vertices. This coefficient is calculated
by making a 4-cut.
The cut-momenta are calculated
explicitly in \cite{Badger:2008cm}. For us, it is only important
that for large internal mass $\mu$, the momentum behaves like:
\begin{equation}
\label{boxcut}
p = p_0(\mu) + \left|\mu\right| \chi,
\end{equation}
where $\chi^2 = 1$ and 
$\underset{\mu \rightarrow \infty}{\lim} {p_0 \over \mu}  = \underset{\mu \rightarrow \infty}{\lim} {p_0 \cdot \chi} = 0 $.

We are concerned with the product of 4 tree-amplitudes, each with two scalars
and an arbitrary number of gluons. The first of these has scalar
momenta $p, p+q_1$. To analyze this tree-amplitude we go to a gauge
where the gauge field satisfies
\begin{equation}
\label{largemass}
\chi \cdot A(q) = 0.
\end{equation}
Now, every propagator comes with a factor of ${1 \over \mu}$ as for
example in the figures in the second and third line (Figs. 3 -- 6) of
Table \ref{domtable}.
\TABLE{
\caption{Dominant diagrams at large $z$}
\label{domtable}\centering
\begin{tabular}{|m{0.45\textwidth}|m{0.45\textwidth}|}
\hline
1. \begin{center} \includegraphics*[height=2.4cm]{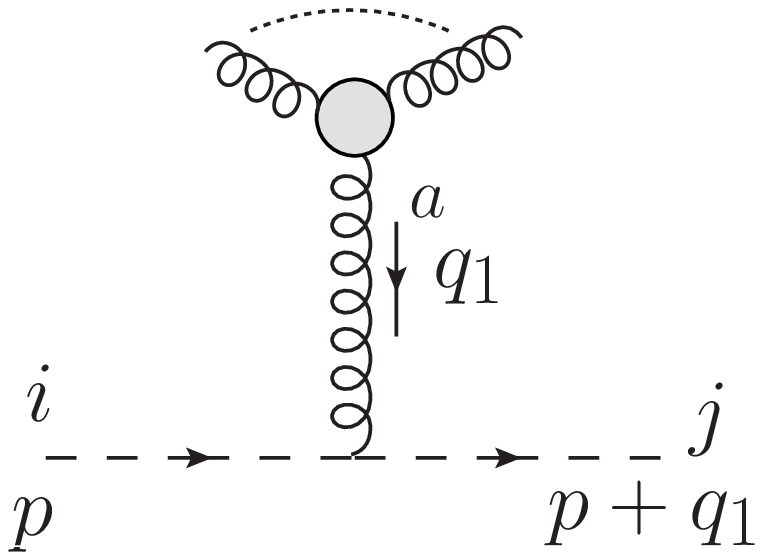} \end{center}&
2.
\begin{center} \includegraphics*[height=2.4cm]{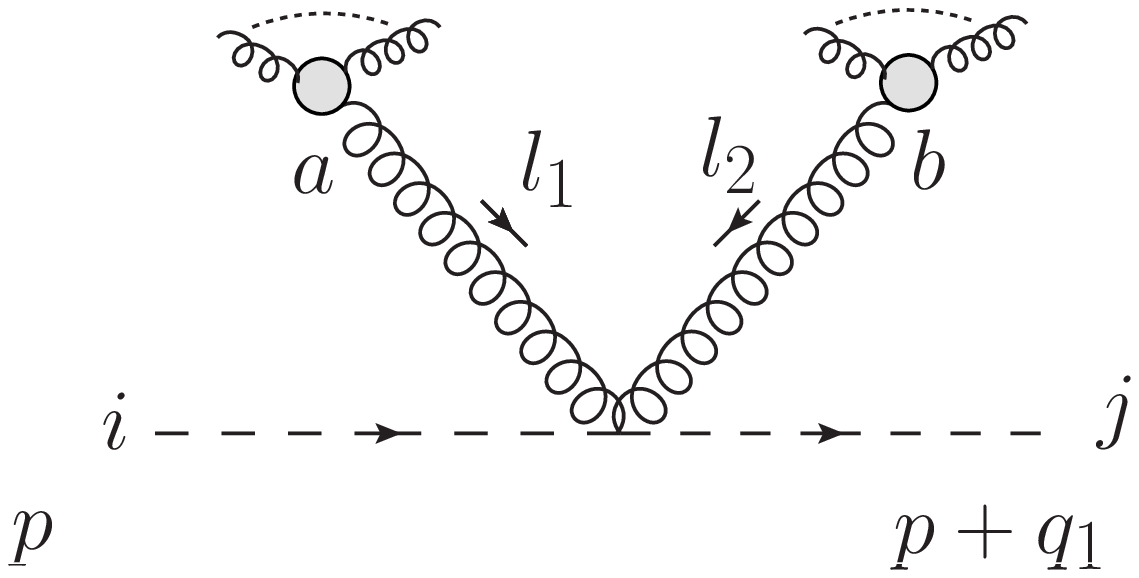} \end{center} \\ \hline
3.
\begin{center} \includegraphics*[height=2.4cm]{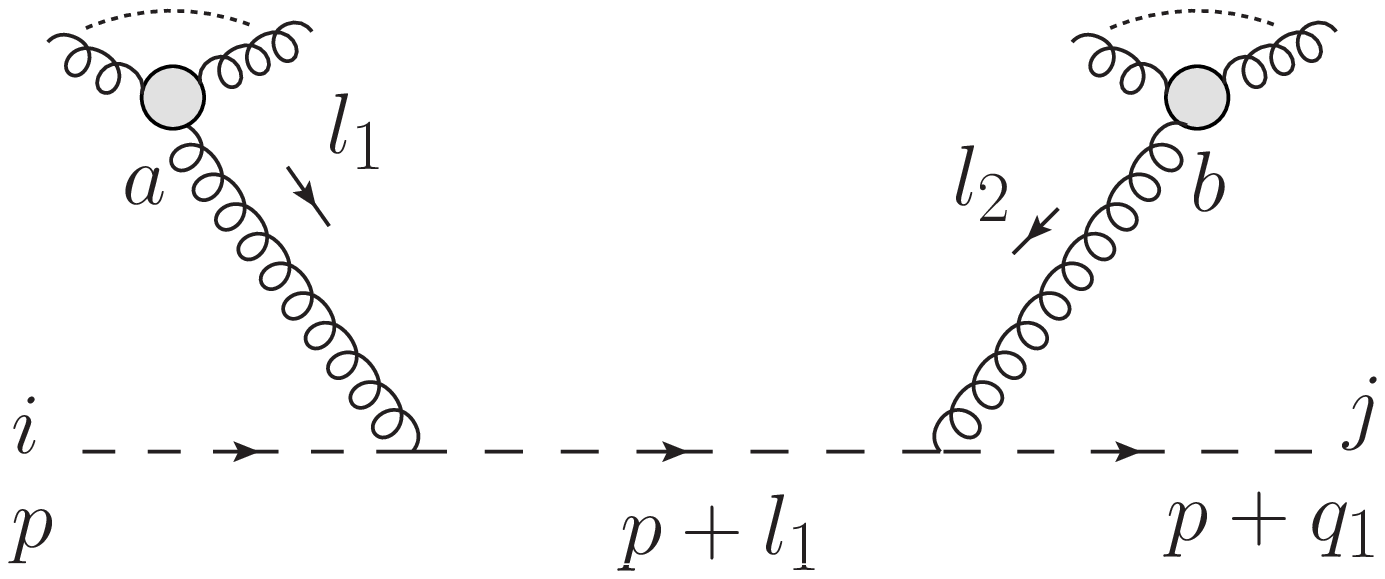} \end{center}&
4.
\begin{center} \includegraphics*[height=2.4cm]{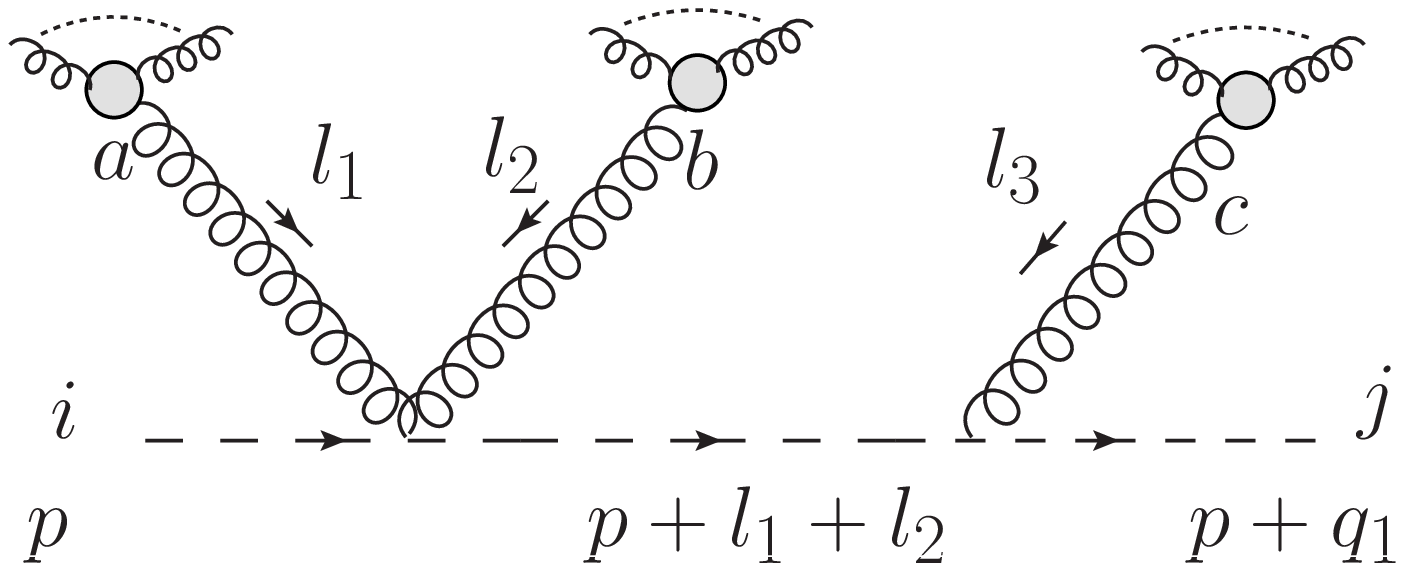} \end{center} \\ \hline
5.
\begin{center} \includegraphics*[height=2.4cm]{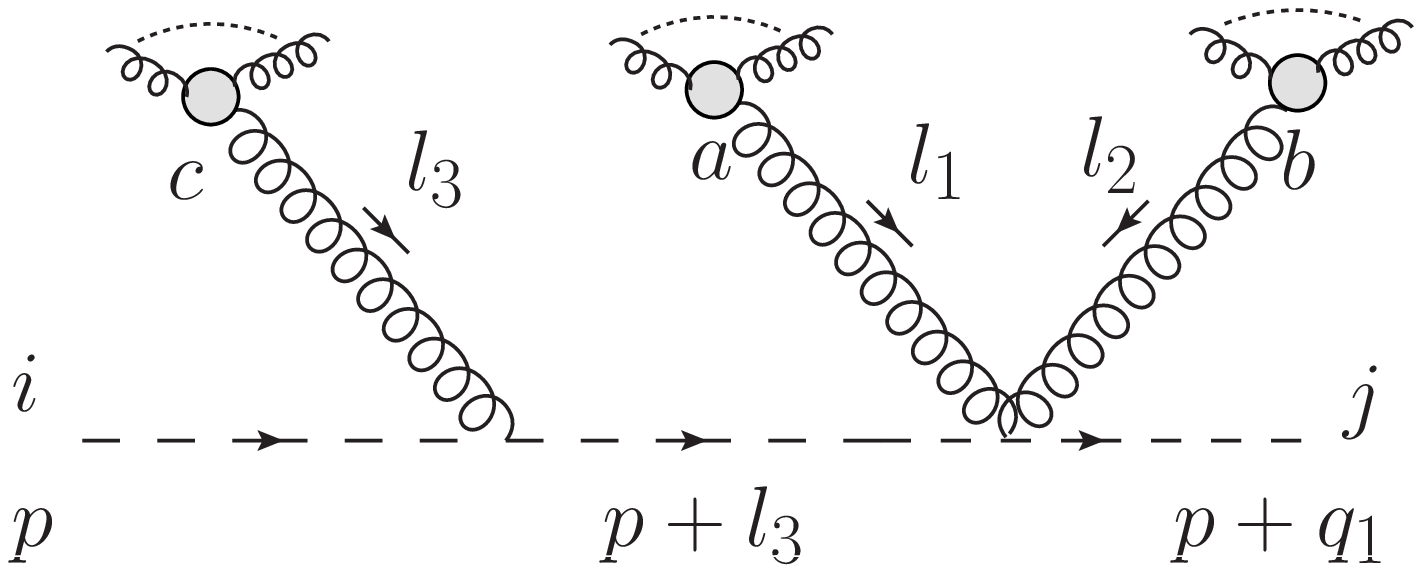} 
\end{center} & 
6.
\begin{center}
\includegraphics*[height=2.4cm]{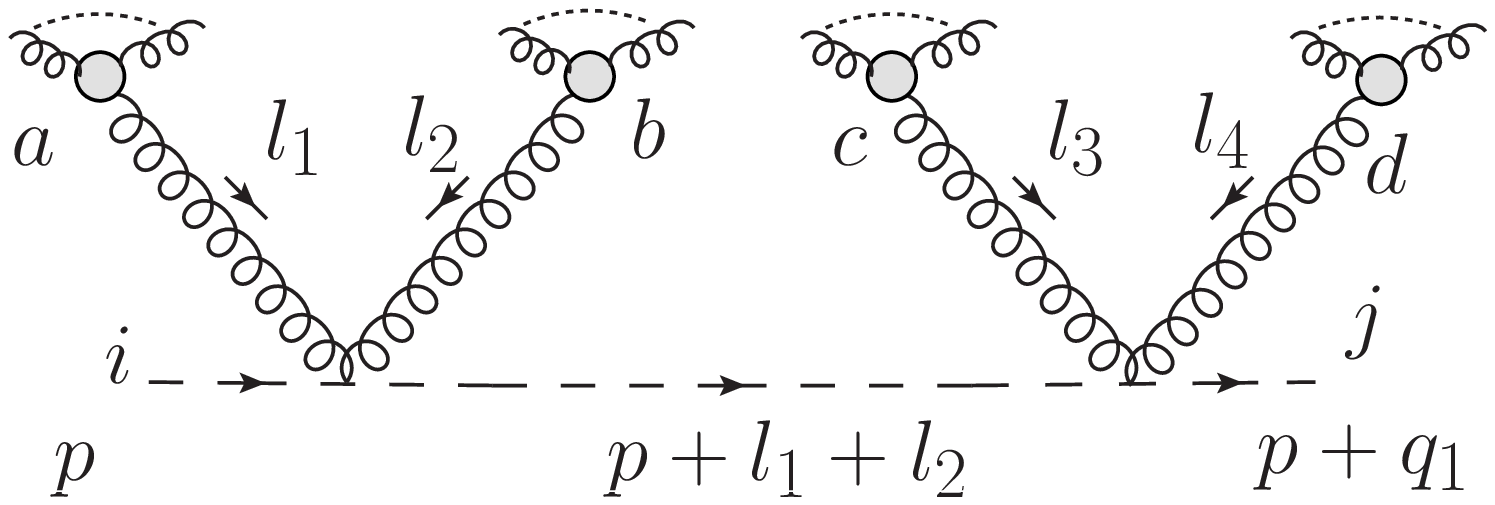} 
\end{center}\\ \hline
\end{tabular}
}
 The only
time we get a factor of $\mu$ in the numerator is when $q = q_1$. 
This
is because $\chi \cdot q_1 = 0$ and so the gauge choice \eqref{largemass}
is not possible. This interaction is shown in the first line (Fig. 1)
of Table \ref{domtable}.
This tells us that the tree-amplitude has the form
\begin{equation}
{\mathcal A}(p, p+q_1) = |\mu| c_a T^a + {c_{a b} \over 2} \{T^{a}, T^{b}\} + \ldots
\end{equation}
where $T^a$ are the generators of the scalar-representation $R$.

In fact we can repeat the analysis of \cite{Lal:2009gn}
to show that, for large $\mu$, the $n$-pt tree-amplitude goes like
\begin{equation}
\label{largemuforbox}
{\mathcal A}(p, p+q_1) = \sum_{k = 1}^n {c_{a_1 \ldots a_k} T^{(a_1}\ldots T^{a_k)} \over |\mu|^{k-2}},
\end{equation}
where $c \underset{\mu \rightarrow \infty}{\longrightarrow} \o[1]$.

In particular, if we want the $\mu^4$ term in the product of four tree-amplitues, we have to take the leading term in the expansion \eqref{largemuforbox}
for each tree-amplitude. Moreover, we need to sum over all scalar colors
to get the box coefficient; this leads to a trace. So, we find that
\begin{equation}
\sum {\mathcal A}_1 {\mathcal A}_2 {\mathcal A}_3 {\mathcal A}_4 = \mu^4 c^1_{a_1} c^2_{a_2} c^3_{a_3} c^4_{a_4} \tr_R\left(T^{a_1} T^{a_2} T^{a_3} T^{a_4}\right) + \o[|\mu|^3].
\end{equation}
The box coefficient is given by further summing this over the two choices
of cut momenta. This implies that the rational contributions from the box-terms can depend
on, at most, the fourth indices of the matter representation. 
More precisely, the rational contribution from the box-coefficient, $R_A$
can be written as
\begin{equation}
\label{boxtorational}
R_{A} = \omega^A_{i_1 i_2} \tr_R \left(T^{i_1} T^{i_2}\right)  + \omega^A_{i_1 i_2 i_3 i_4} \tr_R \left(T^{(i_1} T^{i_2} T^{i_3} T^{i_4)}\right).
\end{equation}
Here, we follow the conventions of \cite{Lal:2009gn} so that a complex scalar in representation $r$ has $R = r + \overline{r}$.
The advantage of this notation is that it makes manifest the fact that
 symmetrized traces of an odd number of generators never appear in the scalar
contribution. (Another way to see this cancellation is to recall that we need to sum over the
two possible orientations of the scalar line in the loop.)

\subsection{Triangles}
Rational terms also come from the $\o[\mu^2]$ term in triangle-coefficients.
To extract triangle coefficients, we make a 3-cut. The 3-cut leaves
us with one free parameter $z$. There are several equivalent ways of 
fixing this parameter and extracting the triangle coefficient \cite{Forde:2007mi,ArkaniHamed:2008gz,Raju:2009yx}. We stick to the conventions of \cite{Forde:2007mi}. As 
the reader
can verify using the detailed formulas in \cite{Badger:2008cm} (we use $z$ instead of $t$), the cut-momentum behaves like 
\begin{equation}
\label{largeztriangles}
p = \chi_0 + z \chi_1 + {\mu^2 \over z} \chi_2.
\end{equation}
What is important for us is that
\begin{equation}
\chi_1^2 = \chi_2^2 = 0; \quad \chi_1 \cdot \chi_2 = 1; \quad \chi_1 \cdot \chi_0 = \o[{1 \over z}]; \quad \chi_2 \cdot \chi_0 = 0.
\end{equation}
The three-cut momenta are $p, p+q_1, p-q_3$. We now wish to take the product of three-amplitudes with two scalars each and several gluons. The rational
term depends on the coefficient of $z^0 \mu^2$ in this product
\begin{equation}
R_{B} = {1 \over 2} 
\sum  \left. {\mathcal A}_1 {\mathcal A}_2 {\mathcal A}_3 \right|_{z^0 \mu^2},
\end{equation}
where the sum is over the intermediate scalar colors and the two solutions for the loop momentum and the 
coefficient is extracted by series expanding first with respect to $z$ around
$z = \infty$ and then with respect to $\mu$ around $\mu = \infty$.

An amplitude with scalar momenta $p, p+q_1$ is dominated by a few diagrams
in the gauge
\begin{equation}
\chi_1 \cdot A(q) = 0.
\end{equation}
The leading-diagram involves a single scalar-gluon interaction as in 
Fig. 1 of Table \ref{domtable}. Other than this, we also need to consider
Figs. 2 and 3 of that table. Finally, there are the
two diagrams shown in Figs. 4 and 5.
Note that each of these would seem to give a contribution to the 
symmetrized product of 3-generators that goes like ${\mu^2 \over z^2}$. 
This combines a ${\mu^2 \over z}$ from the 3-pt vertex with
a ${1 \over z}$ from the propagator. However, notice that this term
exactly cancels between the two diagrams. The diagram in Fig. 6 contributes
to a 3-generator term (the symmetrized 4-generator term cancels with
a `flipped' diagram) without a $\mu^2$. Adding all these
contributions, we find that
the behaviour at large $z$ and large $\mu$ 
of a tree-amplitude is 
\begin{equation}
\label{largeztreefirstfew}
A = (a_i z + b_i {\mu^2 \over z})T^i + (a_{i_1 i_2} + b_{i_1 i_2} {\mu^2 \over z^2})T^{(i_1}T^{i_2)} + ({a_{i_1 i_2 i_3} \over z} + {b_{i_1 i_2 i_3} \mu^2 \over z^3}) T^{(i_2}T^{i_2}T^{i_3)} + \o[{1 \over \mu}].
\end{equation}

From here it directly follows that the rational contribution from 
triangles which comes from the $z^0 \mu^2$ term in the product of three-amplitudes must go like
\begin{equation}
\label{triangletorational}
R_{B} = \omega^B_{i_1 i_2} \tr_R \left(T^{i_1} T^{i_2}\right)  + \omega^B_{i_1 i_2 i_3 i_4} \tr_R \left(T^{(i_1} T^{i_2} T^{i_3} T^{i_4)}\right),
\end{equation}
and so, can depend, at most on the fourth index.

\subsection{Bubbles}
The contribution of bubble coefficients to the rational remainder is again
obtained by extracting the $\o[\mu^2]$ piece of the bubble 
coefficient at large $\mu$. The bubble coefficient is extracted
from the two-cut which now leaves two parameters free. An analysis
very similar to the analysis above shows that the rational 
contribution from bubbles can only depend on the quadratic Index.
\begin{equation}
\label{bubbletorational}
R_C = \omega^C_{i_1 i_2} \tr_R\left(T^{i_1} T^{i_2}\right) =  \omega^C_{i_1 i_2} I_2(R) \kappa^{i_1 i_2}.
\end{equation}
This may be seen by parameterizing the two-cut in the form given in \cite{Badger:2008cm,Forde:2007mi} but perhaps the easiest way to see this result is
to use the method of \cite{Raju:2009yx}. Here, the two-cut is parameterized
by putting additional restrictions on a momentum of the form \eqref{largeztriangles}. There are two contributions to the bubble-coefficient; one depends on
the $\mu^2 z^2$ term in the product of three-amplitudes and another depends
on the $\mu^2 z^i, i = 0 \ldots 2$ term in the product of two tree-amplitudes.
Given the behaviour of the tree-amplitude \eqref{largeztreefirstfew}, 
we can see that these terms can depend on, at most, the quadratic index.

\section{Cut-Constructible Theories}
\label{secdiophantine}
From here, we see that it is quite easy to find new cut constructible theories. 
This is because, we just need to satisfy the equation
\begin{equation}
\label{cutconstructible}
\tr_{R_f}\left(T^{(a_1} \ldots T^{a_n)}\right) - {1 \over 2} \tr_{R_s}\left(T^{(a_1} \ldots T^{a_n)}\right) =  \tr_{\rm adj}\left(T^{(a_1} \ldots T^{a_n)}\right), n =2,4.
\end{equation}
Recall that in \eqref{cutconstructible}, we follow the conventions of \cite{Lal:2009gn}, which are explained below \eqref{boxtorational}, and count in terms of {\em real} scalars and {\em Weyl}-fermions.
 Note that for $n=3$, the fermionic trace must vanish for
anomaly cancellation.

The reader might worry that \eqref{cutconstructible} leads to a very large
number of independent equations. For example, for $SU(N)$, one might be led
to believe that \eqref{cutconstructible} consists of $\o[N^8]$ independent
equations corresponding to distinct choices of generators.

In fact, \eqref{cutconstructible} is very simple and leads to just {\em three}
independent equations.\footnote{For the group $SO(8)$, there are {\em four}
independent equations. This is because there are two independent 
invariant tensors of rank $4$. Equation \eqref{expandtraces} must 
also be suitably modified. } This is because the symmetrized trace of two and 
four generators can be expanded as
\begin{equation}
\label{expandtraces}
\begin{split}
&{1 \over 2 } \tr_{R}\{T^{a_1}, T^{a_2}\} = I_2(R) \kappa^{a_1 a_2}, \\
&\tr_R\left[T^{(a_1}T^{a_2}T^{a_3}T^{a_4)}\right] = I_4(R) d^{a_1 a_2 a_3 a_4} +
I_{2,2} \kappa^{(a_1,a_2}\kappa^{a_3 a_4)}.
\end{split}
\end{equation}
The symmetrized trace of 3-generators never appears. For complex scalars and
Dirac fermions this trace cancels when the different contributions to a 
cut are summed over. For real or pseudoreal representations
this trace is zero while for Weyl fermions in complex representations, this
trace must vanish by anomaly cancellation.

Now, expanding the scalar and fermionic representations in terms of 
irreducible representations as
\begin{equation}
R_f = \oplus n_i^f R_i^f, \quad R_f = \oplus n_i^s R_i^s,
\end{equation}

we find \eqref{cutconstructible} can be written as the three-equations
\begin{equation}
\label{cutdiophantine1}
\begin{split}
\sum_{i} \left(n_{i}^f I_{2}(R_i^f) - {n_{i}^s \over 2} I_{2}(R_i^s)\right) &= I_{2}({\rm adj}). \\
\sum_{i} \left(n_{i}^f I_{2,2}(R_i^f) - {n_{i}^s  \over 2}I_{2,2}(R_i^s)\right) &= I_{2,2}({\rm adj}). \\
\sum_{i} \left(n_{i}^f I_{4}(R_i^f) - {n_{i}^s \over 2} I_{4}(R_i^s)\right) &= I_{4}({\rm adj}).
\end{split}
\end{equation}
For the exceptional groups and also for $SU(2), SU(3)$, there are only
two independent equations since $I_{4}$ vanishes. For $SO(8)$, there
is an additional equation since there are two independent fourth-order indices.

The equations \eqref{cutdiophantine1} are a set of linear Diophantine equations in the variables
$n_i$. In fact, given {\em any } solution to the equation
\begin{equation}
\label{cutdiophantine2}
\sum_i n_i I_{\alpha}(R_i) = 0, \quad  n_i \in {\mathcal Z},
\end{equation}
where $I_{\alpha}$ runs over the set $I_2, I_{2,2}, I_{4}$, 
subject to the conditions
\begin{equation}
\label{condition}
 \sum_{n_i > 0} n_i I_{3}(R_i) = 0, 
\end{equation}
we can construct a valid solution to \eqref{cutdiophantine1} by 
taking 
\begin{equation}
\label{onetotheother}
R_f = \text{adj} + \sum_{n_i > 0} n_i R_i, \quad R_s = \sum_{n_i < 0} n_i (R_i + \overline{R_i}).
\end{equation}
The condition \eqref{condition} just imposes that the fermionic representation be anomaly-free.

This implies that in fact \eqref{cutdiophantine2} has an {\em infinite}
number of solutions. This is because  \eqref{cutdiophantine2} leads to 
an underdetermined set of linear Diophantine equations in 
integer variables with {\em rational}
coefficients (since all indices must be rational numbers). This has an infinity of solutions.

This is in sharp contrast to what happens
for one-loop integral coefficients. The equations that result from
Table \ref{simpleconds} have only a finite (and small) number of solutions.
This is because the $n_i$ there are constrained
to be {\em natural numbers} and unlike in \eqref{cutconstructible} the
conditions of Table \ref{simpleconds} have no minus sign. This positivity
constraint is what makes finding solutions hard. In contrast 
\eqref{cutdiophantine2} leads to 
an underdetermined set of linear Diophantine equations in 
{\em integer} variables and as we argued above this has an infinite number of solutions.

Note that the supersymmetric ``next-to-simplest'' theories considered in \cite{Lal:2009gn}
are automatically free of rational terms by the argument in the beginning of 
section \ref{secrationalmatter}. It is easy to see that the non-supersymmetric
theories considered in \cite{Lal:2009gn} all satisfy \eqref{cutdiophantine2}.
This is because any set of 
representations $R_s$ and $R_f$ satisfying the conditions of Table \ref{simpleconds} automatically satisfy
\eqref{cutdiophantine1}.

\subsection{Examples}
Detailed formulae for $I_4, I_{2,2}, I_2$ are given in \cite{okubo:8}. 
We reproduce these formulae in the Appendix. In Table \ref{soltable}, 
we list some solutions to \eqref{cutdiophantine2} for the first
few $SU(N)$ groups. A superscript\textsuperscript{*}
means that the conjugate representation appears with the same multiplicity
\TABLE{
\label{soltable}
\caption{Simple Solutions to \eqref{cutdiophantine2}. }
\begin{tabular}{|c|c|}
\hline Group&Representations \\\hline
SU(2)&-3[1]+2[2]-5[3]+4[4]-[5]\\\hline
SU(3)&-[2,0]\textsuperscript{*}+[2,1]\textsuperscript{*}-[3,0]\textsuperscript{*}\\\hline
SU(4)&[1,0,0]\textsuperscript{*}-3[2,0,0]\textsuperscript{*}+3[1,1,0]\textsuperscript{*}-6[0,1,0] - 2[0,2,0]\\\hline
SU(5)&10[1,0,0,0]\textsuperscript{*}+3[2,0,0,0]\textsuperscript{*}--3[1,1,0,0]\textsuperscript{*}+[0,2,0,0]\textsuperscript{*}\\\hline
\end{tabular}
}
As mentioned above,
these solutions can immediately be converted into a solution to \eqref{cutdiophantine1} using \eqref{onetotheother}. For example, for the solution in the second line of Table \ref{soltable}, we can take
\begin{equation}
\label{examplesol}
R_s = 2\left([2,0]+ [0,2] + [3,0] + [0,3]\right), \quad R_f = [2,1]+[1,2]+[1,1].
\end{equation}
The reader can easily construct other solutions using Appendix \ref{technicalappendix}.

\section{Conclusions}
\label{secconclusions}
We considered rational terms associated with gluon amplitudes in gauge 
theories coupled to matter in arbitrary representations.

It has been known for a long time that supersymmetric theories are cut-constructible. We found that, for non-supersymmetric theories, these rational terms were proportional to the second and fourth order indices of the matter representation. This is summarized in \eqref{boxtorational}, \eqref{triangletorational}, \eqref{bubbletorational}. This led to the conclusion that gluon
amplitudes in a theory would be cut-constructible if \eqref{cutdiophantine1}
was satisfied. Alternately, given any solution to \eqref{cutdiophantine2},
we can construct a solution to \eqref{cutdiophantine1} by means of \eqref{onetotheother}. 

We showed that all the ``next-to-simplest'' quantum field theories of 
\cite{Lal:2009gn} satisfied this relation; moreover, there are an 
infinite number of solutions to \eqref{cutdiophantine2} (and consequently to \eqref{cutdiophantine1}) some of which
are enumerated in Table \ref{soltable}.

This study provides new examples of theories that,
by naive power-counting, are not cut-constructible but in which rational
terms do, in fact, vanish for amplitudes involving gluons. It would be interesting to understand this
directly from Feynman diagrams. Second, rational terms are often a 
complication in the calculation of higher-loop amplitudes. This study
indicates that these computations would simplify for the cut-constructrible
theories discussed here. This should help in developing 
extensions of S-matrix techniques to higher orders in 
perturbation theory. In fact it would be very interesting to understand if 
the simplifications described above persist to higher loops and also to 
amplitudes involving external matter particles for at least some of the 
theories discussed here.

\section*{Acknowledgements}
We would like to thank Zvi Bern, Lance J. Dixon, Rajesh Gopakumar and Anirbit Mukherjee  for 
helpful discussions.

\appendix
\section*{Appendix}
\section{Formulae for indices}
\label{technicalappendix}
 For a brief review of Indices we refer the reader to \cite{Lal:2009gn} or to the original work by Okubo and Patera \cite{okubo:219,okubo:2722,patera:1972,okubo:8,Okubo:1978qe,okubo:2382} and also some recent work \cite{vanRitbergen:1998pn}. 

In this appendix, we reproduce the formulae for fourth order indices from \cite{okubo:8}. The basic formula we need is that if we write
\begin{equation}
X = \zeta_{a} T^a,
\end{equation}
where $T^a$ are the generators of the algebra, then 
\begin{equation}
\label{quartic}
\tr_R(X^4) =  \zeta_{a_1} \zeta_{a_2} \zeta_{a_3} \zeta_{a_4}  \left[{I_4(R) \over I_4({\rm ad})} d^{a_1 a_2 a_3 a_4} + {d_{ad} I_2(R)^2 \over 2 (2 + d_{ad}) d_{R}} \left(6 - 
{c_2^{ad} \over c_2^{R}} \right) \kappa^{a_1 a_2} \kappa^{a_3 a_4}\right],
\end{equation}
where $\kappa$ is the Killing form, $d_R$ is the dimension of representation
$R$ and $c_2$ is the second Casimir. This relation is valid for all algebras
except for $SO(8)$; the reader may consult \cite{okubo:8} for this special
case. 

We work in the orthogonal basis (see \cite{Lal:2009gn} for 
the relation between the orthogonal and the Dynkin bases) with the highest-weights denoted by $o_i$. Furthermore, with $\rho_i$ the half-sum of positive weights, we define
\begin{equation}
\sigma_i = o_i + \rho_i.
\end{equation}

In each case, the dimension may be calculated by the Weyl dimension formula (see page 233 of \cite{fuchs1997sla}). Moreover, $I_{2,2}$ can be read off from 
\eqref{quartic} and  
\begin{equation}
I_{2,2}(R) = {d_{ad} I_2(R)^2 \over 2 (2 + d_{ad}) d_{R}} \left(6 - 
{c_2^{ad} \over c_2^{R}} \right).
\end{equation}
Note that $I_2(R)^2  \propto {I_{2}(R)^2 \over d_R} \propto d_R (c_2^R)^2$ where $c_2^R$ is the quadratic Casimir. Finally, we have the following formulae for
$I_2(R)$ and $I_4(R)$.
\paragraph{$A_{n-1}$}
\begin{equation}
\begin{split}
{I_2(R) \over d_R} &= \sum_{j=1}^n \sigma_j^2 - {n (n^2 - 1) \over 12}, \\
{I_4(R) \over d_R} &= (n^2 + 1) \sum_{j=1}^n \sigma_j^4 - {2 n^2 - 3 \over n} \left[\sum_{j=1}^n \sigma_j^2\right]^2 + {1 \over 720} n(n^2 - 1)(n^2 - 4)(n^2 - 9).
\end{split}
\end{equation}
\paragraph{$B_n$}
\begin{equation}
\begin{split}
{2 I_2(R) \over d_R} &= \sum_{j=1}^n \sigma_j^2 - {n (4 n^2 - 1) \over 12},\\
{8 I_4(R) \over d_R} &=  (2 n^2 + n + 2) \sum_{j=1}^n \sigma_j^4 - (4 n + 1)  \left[\sum_{j=1}^n \sigma_j^2\right]^2  + {1 \over 360} n (n^2 - 1)(4 n^2 - 1) (2 n + 3) (2 n -7).
\end{split}
\end{equation}
\paragraph{$C_n$}
\begin{equation}
\begin{split}
{2 I_2(R) \over d_R} &= \sum_{j=1}^n \sigma_j^2 - {n (n+1)(2 n + 1) \over 6},\\
{8 I_4(R) \over d_R} &= (2 n^2 + n + 2) \sum_{j=1}^n \sigma_j^4 - (4 n + 1)  \left[\sum_{j=1}^n \sigma_j^2\right]^2  + {1 \over 180} n (n^2 - 1)(4 n^2 - 1) (2 n + 3) (n +4).
\end{split}
\end{equation}
\paragraph{$D_n$}
\begin{equation}
\begin{split}
{2 I_2(R) \over d_R} &= \sum_{j=1}^n \sigma_j^2 - {n (n-1)(2 n - 1) \over 6},\\
{8 I_4(R) \over d_R} &=  (2 n^2 - n + 2) \sum_{j=1}^n \sigma_j^4 - (4 n - 1)  \left[\sum_{j=1}^n \sigma_j^2\right]^2  + {1 \over 180} n (n^2 - 1)(4 n^2 - 1) (2 n - 3) (n -4).
\end{split}
\end{equation}
 
\bibliographystyle{JHEP}
\bibliography{references}
\end{document}